\documentstyle[12pt]{article}
\begin{document}
\title{Relativistic Shifted - $l$ Expansion Technique for Dirac and
 Klein-Gordon Equations}
\author{ Omar Mustafa \\
 Department of Physics, Eastern Mediterranean University\\
 G. Magusa, North Cyprus, Mersin 10 - Turkey\\
 Thabit Barakat \\
 Department of Civil Engineering, Near East University\\
 Lefko{\c s}a, North Cyprus, Mersin 10 - Turkey\\
\date{}\\}
\maketitle
\begin{abstract}
{\small The shifted - $l$ expansion technique (SLET) is
extended to solve for Dirac particle trapped in spherically symmetric
scalar and/or 4-vector potentials. A parameter $\lambda=0, 1$ is introduced in
such a way that one can obtain the Klein-Gordon (KG) bound states from
Dirac bound states. The 4-vector Coulomb, the scalar linear, and the equally
mixed scalar and 4-vector
power-law potentials are used in KG and Dirac equations. Exact
numerical results are obtained for the 4-vector Coulomb potential in both
KG and Dirac equations. Highly accurate and fast converging results are
obtained for the scalar linear and the equally mixed scalar and 4-vector
power-law potentials.}
\end{abstract}
\newpage
\renewcommand{\thesection}{\Roman{section}}

\section{Introduction}

 In a previous paper we have developed the shifted-$l$ expansion technique
(SLET) to solve $3D$ and $2D$ Schr\"odinger equations [1]. Therein, we have 
suggested SLET as a reformation to the shifted large-N expansion technique
(SLNT) [2,3].
The importance of the Klein-Gordon (KG) and Dirac wave equations with
scalar and/or 4-vector potentials arises in many fields of physics
[4-24]. In the framework of the mentioned equations, a scalar or an equally
mixed scalar and 4-vector potentials have considerable interest in the
study of quarkonium systems [6,8-10,12,16-19]. The mixture of 4-vector and 
scalar potentials has been studied by Long and Robson [20]. The
4-vector potentials have great utility in atomic, nuclear, and plasma physics
[21-24]. Therefore, many attempts have been made to develop approximation
techniques to treat relativistic particles in the KG and Dirac wave equations
 [4-24].

To the best of our knowledge, the attempts that have been made to develop
a relativistic $1/N$ expansion technique include, Neito [14] who has
 extended the unshifted $1/N$ expansion [25] formalism to treat a
 $\pi$-mesonic atom in the KG equation. Miramontes and Pajares [15]
who have studied the same technique in the framework of the KG and
 Dirac equations. Iterative procedures concerning the $1/N$ expansion
technique were introduced to deal with KG  and Dirac particles [11]. Stepanove
and Tutik [10] who have used the $\hbar$-expansion procedure in a similar
manner to the $1/N$ expansion procedure. Such procedures are straight forward
though rather tedious and suffer from convergence problems in comparison to 
SLNT [4-8]. In addition, highly excited states pose problems for the above
mentioned techniques. Recently, Roychoudhury and Varshni have developed 
SLNT to treat scalar [12] and 4-vector potentials [13] in the Dirac equation.
Panja et al [4] have applied SLNT to Dirac particle by considering the KG 
equation in which a spin-orbit interaction is introduced analogous to Pauli
theory ( this has been proposed by Papp [8]).
 Mustafa and Sever [6-8] have used a different approach to SLNT for
the KG and Dirac equations.

Encouraged by the success of SLET in the Schr\"odinger equation and the 
importance of KG and Dirac equations with scalar and/or 4-vector potentials,
we feel tempted to extend SLET to solve for the bound states of the above
equations. We shall do so by considering the Dirac equation in which we
introduce a parameter $\lambda=0,1$ in the spin-dependent term. For
$\lambda=0$,
 Dirac equation implies KG equation with scalar and/or 4-vector  potentials.
In  this  case, the total angular momentum $j$ is given through the 
relation  $l=j\pm1/2$, and the radial wave function of KG particle
is  the  radial  large-component of the Dirac spinor. When $\lambda=1$, the 
following are the bound states under consideration; (i) the Dirac
bound states in scalar and/or 4-vector potentials, and (ii) the Dirac bound
states in an equally mixed scalar and 4-vector potentials. The
bound states in (ii) are identical to KG bound states in the same structure
potentials.

For the sake of numerical illustration on the accuracy of
our SLET we discuss the results for the following particles; (i) Dirac
 and KG particles in 4-vector
Coulomb potentials, (ii) Dirac and KG particles in scalar linear potentials,
and (iii) Dirac particle in an equally mixed scalar and 4-vector
power-low potential. Elsewhere, we shall investigate Dirac and KG particles
in a mixture of 4-vector and scalar potentials.

In Sec II we shall extend SLET to solve for Dirac and KG bound states.
Analytical expressions are given in such away that allows the reader to use
them without proceeding into their derivations. In Sec III we shall show that 
the analytical expressions of Sec II lead to highly accurate and fast
converging numerical results when applied to the 4-vector Coulomb, the linear
scalar, and the equally mixed scalar and 4-vector power-law potentials in
both Dirac and KG equations. We conclude and remark in Sec IV.

\section{ SLET for Dirac and KG bound states.}

With a scalar $S(r)$ and a 4-vector $V(r)$ Lorentz interactions,
 simultaneously present, the radial coupled Dirac equations can be written
  [16,26] $(\hbar=c=1)$ as \\
 \begin{equation}
\xi_{1}G(r)+\frac{dF(r)}{dr}-\frac{k}{r}F(r)=0,
\end{equation}
\begin{equation}
\xi_{2}F(r)-\frac{dG(r)}{dr}-\frac{k}{r}G(r)=0.
\end{equation}
Where $\xi_{1}=E-m-V(r)-S(r)$, $\xi_{2}=E+m-y(r)$, $y(r)=V(r)-S(r)$, and
$E$ is the relativistic energy. $G(r)$ and $F(r)$ are the large and small
radial components of the Dirac spinor [19], respectively. Provided $V(r)$ and
$S(r)$ are spherically symmetric, $k=-(l+1)$ for the total angular momentum
$j=l+1/2$, $k=l$ for $j=l-1/2$, and $l$ is angular momentum quantum
number.

In terms of the large-component $G(r)$ of the Dirac spinor, Eq.(1) reads \\
\begin{equation}
\left[\frac{d^{2}}{dr^{2}}-\frac{k(k+1)}{r^{2}}+\frac{1}{\xi_{2}}(y^{'}(r)
[\frac{d}{dr}+\frac{k}{r}])+\xi_{1}\xi_{2}\right]G(r)=0,
\end{equation}
where the prime denotes $d/dr$.

For any $k$ one can show that $k(k+1)=l(l+1)$. Furthermore, we remove
the first derivative by proposing the ansatz
\begin{equation}
G(r)=R(r)exp(-p(r)/2)~~~~;~~~~p^{'}(r)=y^{'}(r)/\xi_{2}.
\end{equation}
Which in turn implies the Schr\"odinger-like radial Dirac equation \\
\begin{equation}
\left[-d^{2}/dr^{2}+l(l+1)/r^{2}+U(r)-\xi_{1}\xi_{2}\right]R(r)=0,
\end{equation}
with \\
\begin{equation}
U(r)=\frac{1}{4}\left[\frac{2y^{''}(r)}{\xi_{2}}-\frac{4kp^{'}(r)}{r}+
3p^{'}(r)^{2}\right].
\end{equation}
      
It should be pointed out that the choice of the ansatz in Eq.(4)
is motivated by requiring the agreement between Eq.(5) and the outcome of
the linear matrix eigenvalue problem, Eq.(1) and (2), after being
transformed into a diagonal Sturm-Liouville eigenvalue system. For more
details on this system the reader may refer to Barut [27].
Moreover, it is evident that Eq.(5) reduces to KG wave equation with scalar
and/or 4-vector potentials provided $U(r)=0$ [6]. We therefore introduce
a parameter $\lambda=0, 1$ in $U(r)$ so that $\lambda=0$ and $\lambda=1$
correspond to KG and Dirac wave equations, respectively. Also, we shall be
interested in problems where the rest energy is large compared to the
binding energy $W=E-m$. Eq.(6) thus becomes \\
\begin{equation}
U(r)=\frac{\lambda}{4m}\left[y^{''}(r)-\frac{2ky^{'}(r)}{r}+
\frac{3y^{'}(r)^{2}}{4m}\right],
\end{equation}
where we have considered\\
\begin{equation}
\frac{1}{\xi_{2}}=\frac{1}{W+2m-y(r)}\simeq\frac{1}{2m}(1-\frac{W-y(r)}
{2m}+\cdots ),
\end{equation}
and the term $(W-y(r))/4m^2$ is small enough to be neglected.

If we shift the angular momentum $l$ through the relation $l=\bar{l}+\beta$
and define\\
\begin{equation}
\gamma(r)=-V(r)^{2}+m(r)^{2}+U(r),
\end{equation}
 where\\
\begin{equation}
m(r)=m+S(r),
\end{equation}
Eq.(5) becomes\\
\begin{equation}
\left[-\frac{d^{2}}{dr^{2}}+\frac{[\bar{l}^{2}+\bar{l}(2\beta+1)+
\beta(\beta +1)]}{r^{2}}+\gamma(r)+2EV(r)\right]R(r)=E^{2}R(r).
\end{equation}
We shall now start the systematic $1/\bar{l}$  expansion by defining\\
\begin{equation}
\gamma(r)=\frac{\bar{l}^{2}}{Q}\left[\gamma(r_{o})+\gamma'(r_{o})r_{o}
 x/\bar{l}^{1/2}
+\gamma''(r_{o})r_{o}^{2}x^{2}/2\bar{l}+\cdots\right],
\end{equation}
\begin{equation}
V(r)=\frac{\bar{l}^{2}}{Q}\left[V(r_{o})+V'(r_{o})r_{o}
 x/\bar{l}^{1/2}
+V''(r_{o})r_{o}^{2}x^{2}/2\bar{l}+\cdots\right],
\end{equation}
\begin{equation}
E=\frac{\bar{l}^{2}}{Q}\left[E_{o}+E_{1}/\bar{l}+E_{2}/
 \bar{l}^{2}+E_{3}/\bar{l}^{3}+\cdots\right].
\end{equation}
where $x=\bar{l}^{1/2}(r-r_{o})/r_{o}$, and $Q$ is to be set equal
to $\bar{l}^{2}$ at the end of the calculations.
With Eqs.(12)-(14) into Eq.(11) one gets\\
\begin{eqnarray}
&&\left[\frac{-d^{2}}{dx^{2}}+(\bar{l}+(2\beta+1)+\frac{\beta(\beta+1)}
{\bar{l}})(1-\frac{2x}{\bar{l}^{1/2}}+\frac{3x^{2}}{\bar{l}}
-\cdots)\right. \nonumber \\
&&\nonumber\\
&&\left.+\frac{r_{o}^{2}\bar{l}}{Q}(\gamma(r_{o})+\frac{\gamma'(r_{o})r_{o}
x}{\bar{l}^{1/2}}+\frac{\gamma''(r_{o})r_{o}^{2}x^{2}}{2\bar{l}}+
\frac{\gamma'''(r_{o})r_{o}^{3}x^{3}}{6\bar{l}^{3/2}}+\cdots)\right.
\nonumber \\
&&\nonumber \\
&&\left.+\frac{2r_{o}^{2}\bar{l}}{Q}(V(r_{o})+\frac{V'(r_{o})r_{o}x }
{\bar{l}^{1/2}} +\cdots)(E_{o}+\frac{E_{1}}{\bar{l}}
+\frac{E_{2}}{\bar{l}^{2}}+\cdots)\right]\phi_{n_{r}}(x)
\nonumber \\
&&\nonumber \\
&&=\mu_{n_{r}}\phi_{n_{r}}(x)
\end{eqnarray}\\
 where\\
\begin{equation}
\mu_{n_{r}}=\frac{r^{2}_{0}\bar{l}}{Q}\left[
E_{o}^{2}+\frac{2E_{o}E_{1}}{\bar{l}}
+\frac{(E_{1}^{2}+2E_{o}E_{2})}{\bar{l}^{2}}+\frac{
2(E_{o}E_{3}+E_{1}E_{2})}{\bar{l}^{3}}
+\cdots\right].
\end{equation}
Eq(15) is a Schr\"odinger-like equation for the one-dimensional anharmonic
oscillator and has been discussed in detail by Imbo et al [2]. We
therefore quote only the resulting eigenvalue of Ref.[2] and write\\
\begin{eqnarray}
\mu_{n_{r}}&=&\bar{l}\left[1+\frac{2r_{o}^{2}V(r_{o})E_{o}}{Q}+
\frac{r_{o}^{2}\gamma(r_{o})}{Q} \right]
\nonumber \\
&&\nonumber \\
&&+\left[(2\beta+1)+\frac{2r_{o}^{2}V(r_{o})E_{1}}{Q}+(n_{r}+\frac{1}{2})w
\right]
\nonumber \\
&&\nonumber\\
&&+\frac{1}{\bar{l}}\left[\beta(\beta+1)+\frac{2r_{o}^{2}V(r_{o})E_{2}}{Q}
+\alpha_{1}\right]
\nonumber \\
&&\nonumber\\
&&+\frac{1}{\bar{l}^{2}}\left[\frac{2r_{o}^{2}V(r_{o})E_{3}}{Q}+\alpha_{2}
\right],
\end{eqnarray}\\
where $\alpha_{1}$  and $\alpha_{2}$  are given in the appendix of Ref. [1],
and the corresponding relativistic $\epsilon^{'}$s and $\delta^{'}$s are given
in the appendix of this paper. If we now compare Eq.(17) with (16), we
obtain\\
\begin{equation}
E_{o}=V(r_{o})+\sqrt{V(r_{o})^{2}+Q/r_{o}^{2}+\gamma(r_{o})},
\end{equation}
\begin{equation}
E_{1}=\frac{Q}{2r_{o}^{2}(E_{o}-V(r_{o}))}\left[2\beta+1+(n_{r}+1/2)w\right],
\end{equation}
\begin{equation}
E_{2}=\frac{Q}{2r_{o}^{2}(E_{o}-V(r_{o}))}\left[\beta(\beta+1)+\alpha_{1}
\right],
\end{equation}
\begin{equation}
E_{3}=\frac{Q}{2r_{o}^{2}(E_{o}-V(r_{o}))}\alpha_{2},
\end{equation}
and\\
\begin{equation}
E_{n_{r}}=E_{o}+\frac{1}{2r_{o}^{2}(E_{o}-V(r_{o}))}\left[\beta(\beta+1)+
\alpha_{1}+\frac{\alpha_{2}}{\bar{l}}\right].
\end{equation}
$r_{o}$ is chosen to be the  minimum of $E_{o}$, i. e.;\\
\begin{equation}
dE_{o}/dr_{o}=0 ~~~~and~~~~d^{2}E_{o}/dr_{o}^{2}>0.
\end{equation} \\
 Hence, $r_{o}$  is obtained through the relation\\
\begin{equation}
2(l-\beta)^{2}=b(r_{o})+\sqrt{b(r_{o})^{2}-4c(r_{o})},
\end{equation}
Where\\
\begin{equation}
b(r_{o})=r_{o}^{3}\left[2V(r_{o})V^{'}(r_{o})+\gamma^{'}(r_{o})+r_{o}V^{'}
(r_{o})^{2}\right],
\end{equation}
\begin{equation}
c(r_{o})=\frac{r_{o}^{6}}{4}\left[\gamma^{'}(r_{o})^{2}+4V(r_{o})V^{'}(r_{o})
\gamma^{'}(r_{o})-4\gamma(r_{o})V^{'}(r_{o})^{2}\right]
\end{equation}
The shifting parameter $\beta$ is determined by requiring
$E_{1}=0$ [1,2]. Therefore,\\
\begin{equation}
\beta=-[1+(n_{r}+1/2)w]/2,
\end{equation}
where\\
\begin{equation}
w=\left[12+\frac{2r_{o}^{4}\gamma^{''}(r_{o})}{Q}+
\frac{4r_{o}^{4}V^{''}(r_{o})E_{o}}{Q}\right]^{1/2}.
\end{equation}

\section{ Applications, numerical results, and discussion.} 
  
In this section we shall consider the 4-vector Coulomb, the scalar linear, and
the  equally-mixed scalar and 4-vector power-law potentials in KG and Dirac
equations.
To obtain the KG and Dirac bound states we set $\lambda=0$ and
$\lambda=1$, respectively. $r_{o}$ is obtained from Eq.(24) through (25)-(29),
and the energy eigenvalues are calculated from Eq.(22). We shall
examine the accuracy and the
convergence of SLET by comparing its outcomes with the exact and numerical
results of other methods.

\subsection{The 4-vector Coulomb potential in KG and Dirac equations
[$V(r)=-\alpha/r$, $ S(r)=0$]}

 A pionic atom in a Coulomb-type 4-vector potential, $V(r)= -\alpha/r$  and
$S(r)=0$,
obeys the KG equation. Where $\alpha$ is the fine structure constant and
$\alpha=1/137.03602$.
The pion mass is $m_{\pi}c^{2}=139.577$ MeV and spin-0. An electron trapped 
in the same
type of potential obeys the Dirac equation. The rest mass of the electron is
$m_{e}c^{2}=0.5110041$ MeV. However, for both particles exact
analytical results for binding
energies exist [26] and hence exact numerical results can be reproduced
and compared with those of SLET.

In Tables 1 and 2 we show the results of SLET compared with the exact results
for KG and Dirac particles, respectively. The results are shown in
such a way that one can easily judge the accuracy and the convergence.
 Our results appear in excellent agreement with the exact
ones. The minimum and maximum percent accuracies found for both particles
are $99.998\%$ and $100.000\%$, respectively.
The convergence is noted to be very fast in the sense that the leading
term $E_{o}$  contributes from $99.996\%$ to $100.000\%$ of the exact binding
energy.

\subsection{ The scalar linear potential in KG and Dirac equations
[$V(r)=0$, and $S(r)=Ar$]}

A pure scalar linear potential, $V(r)=0$ and $S(r)=Ar$, in KG and Dirac
equations is precisely a quark confining linear potential. This potential
has been studied by Gunion and Li [17] in KG and Dirac equations and
hence accurate numerical results exist for comparison with SLET results.
However, Mustafa et al [6] and Roy et al [12] have used SLNT to calculate
part of the KG and Dirac mass spectra, respectively, of some quark-
antiquark systems.

In Tables 3-5 we compare SLET results with those of SLNT [6,12] and those
of Gunion and Li [17]. Comparing our results, in Table 3, with the 
numerically predicted
ones [17] we have scored a minimum accuracy of $99.96\%$ and a maximum
accuracy of $100.00\%$. In Table 4 the minimum accuracy is noted to
be $99.61\%$ and the maximum is $99.76\%$. In Table 5 the accuracy ranges
from $99.05\%$ to $100.00\%$. Furthermore, the energy eigenvalues given
by Eq.(14)
have been noted to be fast converging in the sense that the leading term
$E_{o}$  contributes more than $99\%$ of the total energy. While our results
in Table 3 are in exact agreement with those of SLNT [6], they are only in
qualitative agreement with those of SLNT [12] in Tables 4 and 5. The term
that appears as $3y^{'}(r)^{2}/8m^{2}$ in Eq. (7) of this text has
appeared as $y^{'}(r)^{2}/8m^{2}$
in Eq.(10a) of Ref.[12]. Their method [12] was restricted
only to the scalar linear potential $S(r)=Ar$ in the Dirac equation
where the effective potential can easily be treated by the usual
nonrelativistic SLNT. 
In addition the authors of Ref.[12] were interested in terms only up to
$O(A/m^{2})$, a thing we have not assumed here. However, the
term $3y^{'}(r)^{2}/8m^{2}$ has been confirmed to exist by Barut [27].
 Moreover, it is noteworthy
to mention that for some values of $n_{r}$ and $l$ the standard numerical
method [17] could not predict the effect of the spin-orbit coupling term
$\vec L.\vec S$. SLET was sensitive enough to predict this effect.
To see this one may compare the results of Table 3 with those of Table 4.

\subsection{ The equally mixed scalar and 4-vector power-law potential
[$V(r)=S(r)=Ar^{\nu}+V_{o}$]}

The power-law  potential of the form\\
\begin{equation}
V(r)=Ar^{\nu}+V_{o}
\end{equation}
with $\nu=0.1$ and $A>0$ is a simple non-QCD- based potential. An equally
mixed scalar and 4-vector structure of this potential, i.e. $V(r)=S(r)$,
 in Dirac equation is known to reproduce the data of $\Psi$ and $\Upsilon$
spectroscopies.

In Tables 6,7,8 and 9 we show the spin-averaged results for the equally mixed
scalar and 4-vector power-law potential, Eq.(29), with $V_{o}=-2.028$ GeV,
$\nu =0.1$, $A=a^{\nu+1}$, a=1.709, $m_{c}=1.6179$ GeV, $m_{b}=5.0114$ GeV,
$m_{s}=0.325$ GeV, and $m_{d}=0.01$ GeV.

In Table 6 we compare the eigenvalues $\varepsilon_{n_{r}l}$ corresponding
to the confined bound states of quarks. $\varepsilon_{n_{r}l}$ is calculated
through the relation [19]\\
\begin{equation}
\varepsilon_{n_{r}l}=(E_{c}-m_{c}-2V_{o})\left[(E_{c}+m_{c})(2A)^{-2/\nu}
\right]^{\nu/(\nu+2)}
\end{equation}
In this table the accuracies of our results are in the range $99.90\%$
to $100.00\%$. Whereas in Ref.[5], SLNT has scored a maximum error
of $2.7\%$ corresponding to a minimum accuracy of $97.3\%$. Therein, the
authors [5]
should have noticed that the term on the right hand side of Eq.(5), in
Ref.[5], vanishes for the equally mixed scalar and 4-vector potential case.
This makes it unnecessary to seek the nonrelativistic limit they
have sought. The consideration of this limit in Ref.[5] has caused a spin-
orbit coupling term to exist (see Eq.(9) of Ref. [5]) although it is well
known that one of the interesting features of such mixture cancels any
spin-orbit coupling effect [18].
In Tables 7, 8 and 9 we investigate the accuracy of SLET for some heavy,
$b\bar{b}$, light, $s\bar{s}$, and non-self-conjugate atom like
$c\bar{d}$ mesons, respectively.
Therein, SLET results are found comparable to the numerically predicted
ones [19].
    
\section{ Conclusions and remarks}

In this paper we have extended the shifted-$l$ expansion technique (SLET)
to solve for the eigenvalues of KG and Dirac equations with scalar and/or
4-vector potentials. We have shown that Dirac equation reduces to KG
equation, provided that the angular momentum quantum number is given through
the relation $l=j\pm 1/2$, and the radial KG wave function is the radial
large-component of the Dirac spinor. A parameter $\lambda=0,1$ has been
introduced in the spin-dependent term $U(r)$ so that when $\lambda=0$
the Dirac equation reduces to the KG equation.
The formalism developed in Sec II has been investigated through the
4-vector Coulomb, the scalar linear, and the equally mixed scalar and
4-vector potentials in Dirac and KG equations. SLET results were
found in excellent agreements with the exact and numerically predicted
results found elsewhere.

To the best of our knowledge, this is the first time that both Dirac
and KG equations, with arbitrary scalar and/or 4-vector potentials, have
been solved through one analytical procedure that yields highly accurate and
fast converging results.

In conclusion, SLET is to be understood as being an expansion through
some existing quantum number, which depends on the symmetry of the
potential of interest, in any Schr\"odinger - like eigenvalue problem.
For example, in the case of spherically symmetric potentials the
expansion parameter is $1/\bar{l}$ where $\bar{l}=l-\beta$ and in the case
of cylindrically symmetric potentials $\bar{l}=|m|-\beta$ where $m$ is
the magnetic quantum number. This understanding
suggests we should not worry about the $N$-dimensional form
of the wave equation in hand and we should expand directly through the
quantum numbers involved in the problem. Therefore, the difficulties
associated with inflating the dimensions of the Dirac equation are
resolved by using SLET. The attendant method seems more flexible than
the SLNT.

Finally, we would like to mention that the formalism we have
introduced in this work is also applicable to more complicated
relativistic potentials. In particular, screened Coulomb potentials
which have great utility in atomic, nuclear, and plasma physics. Also, the
wave functions, and normalizations for spin-1/2 and spin-0 particles
in spherically symmetric scalar and/or 4-vector potentials can be
obtained.
\begin{center}
{\bf Appendix}
\end{center}

The definition for  $\varepsilon_{j}'s$ and
$\delta_{i}'s$  in $\alpha_{1}$ and $\alpha_{2}$ of Ref.[1] are as follows
\begin{equation}
\varepsilon_{1}=-2(2\beta+1) {~~,~~}
\varepsilon_{2}=3(2\beta+1),
\end{equation}
\begin{equation}
\varepsilon_{3}=-4+\frac{r_{o}^{5}}{6Q}\left[\gamma^{'''}(r_{o})+2V^{'''}
(r_{o})E_{o}\right],
\end{equation}
\begin{equation}
\varepsilon_{4}=5+\frac{r_{o}^{6}}{24Q}\left[\gamma^{''''}(r_{o})+2V^{''''}
(r_{o})E_{o}\right],
\end{equation}
\begin{equation}
\delta_{1}=-2\beta(\beta+1)+\frac{2r_{o}^{3}V^{'}(r_{o})E_{2}}{Q}
\end{equation}
\begin{equation}
\delta_{2}=3\beta(\beta+1)+\frac{r_{o}^{4}V^{''}(r_{o})E_{2}}{Q},
\end{equation}
\begin{equation}
\delta_{3}=-4(2\beta+1) {~~,~~}
\delta_{4}=5(2\beta+1),
\end{equation}
\begin{equation}
\delta_{5}=-6+\frac{r_{o}^{7}}{120Q}\left[\gamma^{'''''}(r_{o})+2V^{'''''}
(r_{o})E_{o}\right],
\end{equation}
\begin{equation}
\delta_{6}=7+\frac{r_{o}^{8}}{720Q}\left[\gamma^{''''''}(r_{o})+2V^{''''''}
(r_{o})E_{o}\right].
\end{equation}

The terms including $E_{1}$ have been
dropped from the expressions above since $E_{1}$=0.

\newpage
\begin{table}
\begin{center}
\caption{Comparison of the energy levels (in keV units) for spin-0
particle in $V(r)=-\alpha/r$ and $S(r)=0$. W represents the binding
energies, where
$W_{o}= (E_{o}-1)m_{\pi}c^{2}$, $W_{2}=(E_{o}+E_{2}/\bar{l}^{2}-1) m_{\pi}c^{2}$,
and $W_{3}=(E_{o}+E_{2}/\bar{l}^{2}+E_{3}/\bar{l}^{3}-1)m_{\pi}c^{2}$.}
\vspace{1cm}
\begin{tabular}{|c|c|c|c|c|}
\hline
State & -$W_{o}$ &-$W_{2}$ &-$W_{3}$&Exact[26] \\
\hline
$1s$  &  3.71649 & 3.71653   & 3.71656  &3.71659  \\
$2s$  &  0.92910  & 0.92910    &  0.92911 &   0.92912  \\
$2P$  &  0.92910  & 0.92910    &  0.92910  & 0.92909  \\
$3s$  &  0.41293  & 0.41293    &  0.41293  &  0.41294  \\
$3P$  &  0.41293  & 0.41293   &  0.41293  &  0.41293    \\
$3d$  &   0.41293  & 0.41293   &  0.41293  &  0.41293\\
$4s$  &0.23227& 0.23227 &0.23227  & 0.23228\\
$4P$  &0.23227& 0.23227 &0.23227  & 0.23227\\
$4d$  &0.23227& 0.23227 &0.23227  & 0.23227\\
$4f$  &0.23227& 0.23227 &0.23227  & 0.23227\\
\hline
\end{tabular}
\end{center}
\end{table}
\newpage
\begin{table}
\begin{center}
\caption{Comparison of the energy levels (in eV units) for spin-1/2
 particle
 in $V(r)=-\alpha/r$ and $S(r)=0$. W represents the binding energies,
 where $W_{o}=(E_{o}-1)m_{e}c^{2}$, $W_{2}=(E_{o}+E_{2}/\bar{l}^{2}-1)m_{e}c^{2}$,
 and $W_{3}=(E_{o}+E_{2}/\bar{l}^{2}+E_{3}/\bar{l}^{3}-1)m_{e}c^{2}$.}
 \vspace{1cm}
\begin{tabular}{|c|c|c|c|c|}
\hline
State & -$W_{o}$ &-$W_{2}$ &-$W_{3}$&Exact[26] \\
\hline
$1s_{1/2}$  &  13.60639 & 13.60657   & 13.60667  &13.60603  \\
$2s_{1/2}$  &  3.40151  & 3.40153    &  3.40155  & 3.40152  \\
$2P_{1/2}$  & 3.40150   & 3.40151    &  3.40151  & 3.40152  \\
$2P_{3/2}$  &  3.40147  & 3.40147    &  3.40147  & 3.40148  \\
$3s_{1/2}$  &  1.51177  & 1.51177    &  1.51178  & 1.51178  \\
$3P_{1/2}$  &  1.51177  & 1.51178    &  1.51177  & 1.51178  \\
$3P_{3/2}$  &   1.51177 &  1.51177   &  1.51177  & 1.51177    \\
$3d_{3/2}$  &  1.51177  &  1.51177   &  1.51177  & 1.51177  \\
$3d_{5/2}$  &   1.51176 &  1.51176   &  1.51176  & 1.51177\\
$4s_{1/2}$  &0.85037& 0.85037 &0.85037  & 0.85038\\
$4P_{1/2}$  &0.85037& 0.85037 &0.85037  & 0.85038\\
$4P_{3/2}$  &0.85037& 0.85037 &0.85037  & 0.85037\\
$4d_{3/2}$  &0.85037& 0.85037 &0.85037  & 0.85037\\
$4d_{5/2}$  &0.85037& 0.85037 &0.85037  & 0.85037\\
$4f_{5/2}$  &0.85037& 0.85037 &0.85037  & 0.85037\\
$4f_{7/2}$  &0.85037& 0.85037 &0.85037  & 0.85037\\
\hline
\end{tabular}
\end{center}
\end{table}
\newpage
\begin{table}
\begin{center}
\caption{KG results for part of the mass spectra (in GeV) of $\Psi$, $\Psi^{'}$
 system with $S(r)=Ar$, $A=0.137 GeV^{2}$ and $m=1.12 GeV$. The values in
 round parentheses are those of SLNT [6]. The values in square parentheses
are those of Ref.[17].}
\vspace{1cm}
\begin{tabular}{ccccc}
\hline
\multicolumn{1}{c}{}&\multicolumn{4}{c}{l}\\ \cline{2-5}
$n_{r}$ & 0 & 1& 2 & 3 \\
 \hline
 0 & 3.12    & 3.46    & 3.74    & 3.99  \\
   & (3.12)  & (3.46)  & (3.74)  & (3.99)\\
   & [3.1]   & [ 3.47] & [3.73]  & [3.98]\\
 1 & 3.7     & 3.95    & 4.18    & 4.387 \\
   & (3.7)   & (3.96)  & (4.18)  & (4.387)\\
   & [3.7]   & [3.95]  & [4.17]  & [4.387]\\
 2 & 4.15    & 4.36    & 4.56    & 4.74  \\
   & (4.16)  & (4.36)  & (4.56)  & (4.74)\\
   & [4.17]  & [4.38]  & [4.56]  & [4.73]\\
 3 & 4.537   & 4.72    & 4.89    & 5.05  \\
   & (4.537) & (4.72)  & (4.89)  & (5.06)\\
   & [4.54]  & [4.72]  & [4.9]   & [5.05]\\
 4 & 4.88    & 5.04    & 5.195   & 5.35  \\
   & (4.9)   & (5.04)  & (5.198) & (5.35)\\
   & [4.89]  & [5.04]  & [5.195] & [5.35]\\
\hline
\end{tabular}
\end{center}
\end{table}
\newpage
\begin{table}
\begin{center}
\caption{Dirac results for part of the $J/\Psi$ mass spectra (in GeV)
with $S(r)=Ar$, $A=0.137$ $GeV^{2}$, m=1.12 GeV, and $j=l+1/2$. The values
in round parentheses are those of SLNT [12]. The values in square
parentheses are those of Ref.[17].}
\vspace{1cm}
\begin{tabular}{ccccc}
\hline
\multicolumn{1}{c}{}&\multicolumn{4}{c}{l}\\ \cline{2-5}
$n_{r}$ & 0 & 1& 2 & 3 \\
 \hline
 0 & 3.094   &  3.432   & 3.712   & 3.958   \\
   & (3.106) & (3.449)  & (3.733) & (3.982) \\
   & [3.103] & [3.442]  & [3.725] & [3.973] \\
 1 & 3.691   &  3.935   & 4.158   & 4.363   \\
   & (3.698) & (3.948)  & (4.175) & (4.383) \\
   & [3.7]   &  [3.946] & [4.170] & [4.377] \\
 2 & 4.146   & 4.348    & 4.538   & 4.718   \\
   & (4.152) & (4.359)  & (4.553) & (4.736) \\
   & [4.158] & [4.36]   & [4.551] & [4.732] \\
 3 & 4.531   & 4.707    & 4.876   & 5.037   \\
   & (4.535) & (4.717)  & (4.889) & (5.054) \\
   & [4.545] & [4.72]   & [4.89]  & [5.053] \\
 4 & 4.870   & 5.03     & 5.183   & 5.331   \\
   & (4.874) & (5.038)  & (5.195) & (5.345) \\
   & [4.886] & [5.043]  & [5.198] & [5.346] \\
\hline
\end{tabular}
\end{center}
\end{table}
\newpage
\begin{table}
\begin{center}
\caption{Dirac results for part of $J/\Psi$ mass spectra (in GeV)  with
$S(r)=Ar$, $A=0.137 GeV^{2}$ and $m=1.12 GeV$, and $j=l-1/2$. The values in
 round parentheses are those of SLNT [12]. The values in square parentheses
are those of Ref.[17].}
\vspace{1cm}
\begin{tabular}{ccccc}
\hline
\multicolumn{1}{c}{}&\multicolumn{4}{c}{l}\\ \cline{2-5}
$n_{r}$ & 1 & 2& 3 & 4 \\
 \hline
 0 &  3.4726      &  3.7611    &  4.0118  & 4.2369  \\
    &  (3.471)      & (3.760)     &  (4.010)  & (4.236)  \\
     &  [3.47]      & [ 3.757]    &  [4.006]    &  [ 4.23]  \\  
 1   &  3.9646      &  4.1961    &   4.4069  &4.6020  \\
    &  (3.954)      &  (4.194)    &   (4.406)  &(4.600)  \\
    &  [3.965]      & [ 4.194]    &  [ 4.403]    &[4.597]  \\
 2  &4.3713        &4.5699       &  4.7552   & 4.9295\\  
    & (4.370)      & (4.568)      &  (4.754)   & (4.920)  \\
    &[ 4.374]      & [4.560]      &  [4.753]     & [4.926]  \\   
 3   &4.7263        & 4.9030      & 5.0700   &5.2290         \\
     &(4.726)      & (4.902)      & (5.069)    &(5.228)         \\
     &[4.731]      &[ 4.95]      &[ 5.07]     &[$\cdots$]         \\
 4  &5.0456        &5.2062       &5.3595    &5.5065         \\
    &(5.045)       &(5.205)       &(5.359)    &(5.507)         \\
     &[5.053]      &[5.21]       &[5.361]      &[5.506]         \\
\hline
\end{tabular}
\end{center}
\end{table}
\newpage
\begin{table}
\begin{center}
\caption{Spin- averaged Dirac eigenvalues $\epsilon_{n_{r}l}$, Eq(30),
for the equally
 mixed scalar and 4-vector power-law potential $V(r)=S(r)=Ar^{\nu}+V_{o}$,
 $V_{o}=-2.028 GeV$, $\nu=0.1$, $A=a^{\nu+1}$, $a=1.709$,
 and $m_{c}=1.6179 GeV$.}
 \vspace{1cm}
\begin{tabular}{ccccc}
\hline
\multicolumn{1}{c}{}&\multicolumn{3}{c}{$c\bar{c}$}\\ \hline
$n_{r}$ & l & Ref.[19]& SLNT [5] & SLET \\
 \hline
0  &  0  &  1.2364   &  1.2400 &   1.2358  \\
1  &  0  & 1.3347   &  1.3400&   1.3343  \\
2  &  0  & 1.3923   &  1.3980&   1.3915  \\
3  &  0  & 1.4335   &  1.4390&   1.4326  \\
4  &  0  & 1.4657   &  1.4710&   1.4648  \\
0  &  1  & 1.3071    &  1.3090 & 1.3072  \\
1  &  1  & 1.3731   &  1.4110 & 1.3730  \\
0  &  2  & 1.3544   &  1.3580&   1.3540  \\
\hline
\end{tabular}
\end{center}
\end{table}
\newpage
\begin{table}
\begin{center}
\caption{ Spin-averaged Dirac mass spectra (in Gev units), $M_{n_{r}l}=2E$, for
 the $b\bar{b}$ system in the equally mixed scalar and 4-vector
 power-law potential of Eq.(29).}
 \vspace{1cm}
\begin{tabular}{cccc}
\hline
$n_{r}$ & l & Ref.[19]& SLET \\
 \hline
0  &  0  &  9.4390   &   9.4347  \\
1  &  0  & 10.0260  &   10.0228  \\
2  &  0  & 10.3694  &   10.3640  \\
3  &  0  & 10.6147  &   10.6084  \\
4  &  0  & 10.8063  &   10.800  \\
0  &  1  & 9.8613   &    9.8612  \\
1  &  1  & 10.2549  &    10.2537  \\
0  &  2  & 10.1435  &   10.1406  \\
\hline
\end{tabular}
\end{center}
\end{table}
\newpage
\begin{table}
\begin{center}
\caption{ Spin-averaged Dirac mass spectra (in Gev units), $M_{n_{r}l}=2E$,
 for the $s\bar{s}$ system in the equally mixed scalar and 4-vector
 power-law potential of Eq.(29).}
 \vspace{1cm}
\begin{tabular}{cccc}
\hline
$n_{r}$ & l & Ref.[19] & SLET \\
 \hline
0  &  0  &  1.0040    & 0.9999  \\
1  &  0  & 1.5532    &   1.5500  \\
2  &  0  & 1.8804    &   1.8752  \\
3  &  0  & 2.1162    &   2.1099  \\
4  &  0  & 2.3012     &   2.2950  \\
0  &  1  & 1.3976    &   1.3974  \\
1  &  1  & 1.7710    &   1.7697  \\
0  &  2  & 1.6647   &   1.6619  \\
\hline
\end{tabular}
\end{center}
\end{table}
\newpage
\begin{table}
\begin{center}
\caption{Spin-averaged Dirac mass spectra (in Gev units), $M_{n_{r}l}=
E_{d}+m_{c}$,
 for the $c\bar{d}$ system in the equally mixed scalar and 4-vector
 power-law potential of Eq.(29).}
 \vspace{1cm}
\begin{tabular}{cccc}
\hline
$n_{r}$ & l & Ref [19]& SLET \\
 \hline
0  &  0  &  1.9730    &   1.9713  \\
1  &  0  & 2.2107   &   2.2093  \\
2  &  0  & 2.3601   &   2.3577  \\
3  &  0  & 2.4699   &   2.4671  \\
4  &  0  & 2.5571   &   2.5542  \\
0  &  1  & 2.1413   &   2.1413  \\
1  &  1  & 2.3097   &   2.3092  \\
0  &  2  & 2.2612   &   2.2599  \\
\hline
\end{tabular}
\end{center}
\end{table}

\begin{thebibliography}{99}
\bibitem {} O. Mustafa and T. Barakat, J. Phys. {\bf A}, (submitted
 for publication) (1995).
\bibitem {} T. Imbo, N. Pagnamenta, and U. Sukhatme, Phys. Rev. {\bf D29},
1669 (1984).
\bibitem {} O. Mustafa, J. Phys.; Condens. Matter {\bf 5}, 1327 (1993).
\bibitem {} M. Panja, R. Dutt and Y. P. Varshni, Phys. Rev. {\bf A42}, 106
(1990); M. Panja, M. Bag, R. Dutt and Y. P. Varshni, Phys. Rev. {\bf
A42}, 1523 (1992).
\bibitem {} B. Roy and R. Roychoudhury, J. Phys. {\bf A23}, 3555 (1990).
\bibitem {} O. Mustafa and R. Sever, Phys. Rev. {\bf A43}, 5787 (1991);
{\bf A44}, 4142 (1991).
\bibitem {} O. Mustafa and R. Sever, J. Quant. Spectrosc. Radiat. Transfer
{\bf 49}, 65 (1993).
\bibitem {} E. Papp, Ann. Phys. (Leipzig) {\bf 48}, 319 (1991).
\bibitem {} E. Papp, Phys. Lett. {\bf B259}, 19, (1991).
\bibitem {} S. Stepanov and R. Tutik,  Phys. Lett. {\bf A163}, 26, (1992).
\bibitem {} A. Chatterjee, J. Math. Phys. {\bf 27}, 2331 (1986);
S. Atag, J. Math. Phys. {\bf 30}, 696 (1989).
\bibitem {} R. Roychoudhury and Y. Varshni, J. Phys. {\bf A20}, L1083 (1987).
\bibitem {} R. Roychoudhury and Y. Varshni, Phys. Rev.  {\bf A39}, 5523 (1989).
\bibitem {} M. M. Nieto, Am. J. Phys. {\bf 47}, 1067 (1979).
\bibitem {} J. L. Miramontes and C. Pajares, Nouvo Cimento {\bf B84},
10 (1984).
\bibitem {} C. Critchfield, J. Math. Phys. {\bf 17}, 261 (1976).
\bibitem {} J. Gunion and L. Li, Phys. Rev. {\bf D12}, 3583 (1975).
\bibitem {} E. Magyari, Phys. Lett.  {\bf B95}, 295 (1980).
\bibitem {} S. Jena and T. Tripati, Phys. Rev. {\bf D28}, 1780 (1983).
\bibitem {} C. Long and D. Robson, Phys. Rev. {\bf D27}, 644 (1983).
\bibitem {} J. Mc Ennan, D. J. Botto and R. H. Pratt, Phys. Rev. {\bf A16},
1768 (1977).
\bibitem {} O. V. Gabriel, S. Chaudhuri and R. H. Pratt, Phys. Rev. {\bf A24},
3088 (1981).
\bibitem {} E. R. Vrscay and H. Hamidian, Phys. Lett. {\bf A130},
141 (1988).
\bibitem {} G. W. Rogers, Phys. Rev. {\bf A30}, 35 (1984); C. K. Au and
Y. Aharonov, ibid. 20, 2245 (1979).
\bibitem {} L. Mlodinow and M. Shatz, J. Math. Phys. {\bf 25}, 943 (1984).
\bibitem {} W. Greiner, Relativistic Quantum Mechanics (Springer-Verlag, Berlin
, Heidelberg, 1990).
\bibitem {} A. O. Barut, J. Math. Phys. {\bf 21}, 568 (1980).
\end{thebibliography}
\end{document}